\documentclass{elsart}

\usepackage[dvips]{graphicx}
\usepackage{amsbsy}

\newcommand{\bk}{{\bf k}}
\newcommand{\bp}{{\bf p}}
\newcommand{\bq}{{\bf q}}
\newcommand{\br}{{\bf r}}
\newcommand{\bG}{{\bf G}}
\newcommand{\btau}{{\boldsymbol \tau}}
\newcommand{\kF}{k_{\mathrm{F}}}
\newcommand{\EF}{E_{\mathrm{F}}}

\newcommand{\iunit}{\mathrm{i}}
\renewcommand{\Re}{\mathop{\mathrm{Re}}\nolimits}
\newcommand{\Tr}{\mathop{\mathrm{Tr}}\nolimits}

\begin{document}

\runauthor{Angilella, Leys, March, and Pucci}
\runtitle{Linear response function ...}
\journal{J. Phys. Chem. Solids}
\volume{64}
\firstpage{413}
\lastpage{418}
\pubyear{2003}

\begin{frontmatter}
\title{Linear response function around a localized impurity in a
   superconductor}
\author[CT]{G. G. N. Angilella,}
\author[RUCA]{F. E. Leys}
\author[Oxford,RUCA]{N. H. March,}
\and
\author[CT]{R. Pucci}
\address[CT]{Dipartimento di Fisica e Astronomia dell'Universit\`a di
   Catania,\\
and Istituto Nazionale per la Fisica della Materia, UdR di Catania,\\
 64, Via S. Sofia, I-95123 Catania, Italy}
\address[RUCA]{Department of Physics, University of Antwerp (RUCA),\\
   Groenenborgerlaan 171, B-2020 Antwerp, Belgium}
\address[Oxford]{Oxford University, Oxford, England}
\begin{abstract}
Imaging the effects of an impurity like Zn in high-$T_c$
   superconductors [see, \emph{e.g.,} S.~H. Pan \emph{et al.,} Nature
   {\bf 61} (2000) 746] has rekindled interest in defect problems in
   the superconducting phase.
This has prompted us here to re-examine the early work of March and
   Murray [Phys. Rev. {\bf 120} (1960) 830] on the linear response
   function in an initially translationally invariant Fermi gas.
In particular, we present corresponding results for a 
   superconductor at zero temperature, both in the $s$- and in the
   $d$-wave case, and mention their direct physical relevance in the
   case when the impurity potential is highly localized.
\end{abstract}
\end{frontmatter}

\section{Introduction}

The unconventional nature of the pairing state in the high-$T_c$
   superconductors (HTCS) has been recently provided with further
   evidence by the observation of the spatial inhomogeneity of the
   electron distribution around an isolated nonmagnetic impurity, such
   as Zn, in cuprate Bi-2212, imaged by scanning tunneling microscopy
   (STM) \cite{Pan:99,Hudson:99,Hudson:01,Pan:01}.
These experimental studies demonstrated the existence of
   impurity-induced quasi-bound states (resonances) near the Fermi
   level.
Such results have been naturally interpreted as fingerprints of the
   momentum-space anisotropy of the order parameter in the HTSC.
In particular, the four-lobed structure of the STM differential
   conductivity observed around an isolated impurity is a clear
   manifestation of the $d$-wave symmetry of the superconducting gap
   in these materials.
More recently, similar results have been reported for
   Nd/Ba-substituted YBCO thin films \cite{Iavarone:02,Salluzzo:01},
   thus lending further support to the $d$-wave scenario for the
   hole-doped high-$T_c$ cuprates. 

The idea that an anisotropic superconducting gap should give rise to
   directly observable spatial features in the tunneling conductance
   near an impurity was suggested by Byers \emph{et al.}
   \cite{Byers:93}, whereas earlier studies \cite{Choi:90} had
   considered perturbations of the order parameter in unconventional
   superconductors to occur within a distance of the order of the
   coherence length $\xi$ around an impurity.
Later, it was shown that an isolated impurity in a $d$-wave
   superconductor produces virtual bound states close to the Fermi
   level, in the nearly unitary limit \cite{Balatsky:95}.
Such a quasi-bound state should appear as a pronounced peak near the
   Fermi level in the
   local density of states (LDOS) corresponding to an impurity site
   \cite{Salkola:96}, as is indeed observed in Bi-2212 \cite{Pan:99}
   and YBCO \cite{Iavarone:02}.

The location and number of such low-energy peaks in the LDOS around an
   impurity could also provide important new insights into the nature of the
   pseudogap state in the underdoped cuprates above the critical
   temperature $T_c$ \cite{Kruis:01,Wang:02}.
Within the preformed pairs scenario, a pseudogap in the normal state
   is related to the presence of fluctuating pairs, still lacking phase
   coherence, characterized by a much shorter coherence length and
   lifetime than superconducting pairs below $T_c$ \cite{Randeria:97-2}.
On the other hand, the $d$-density wave (DDW) scenario has been
   recently proposed \cite{Chakravarty:01}, in which the normal state
   is characterized by an actual gap in the quasiparticle spectrum,
   due to a hidden broken symmetry of $d_{x^2 - y^2}$-type.
In the presence of a pseudogap depleting the DOS at the Fermi level,
   an impurity-induced resonant state should survive above $T_c$,
   although broadened in energy.
However, if preformed pairs above $T_c$ have a superconducting origin,
   they would be composed of a superposition of hole as well as
   electron states, like Bogoliubov quasiparticles below $T_c$.
Therefore, impurity-induced resonances in the LDOS probed by STM
   spectroscopy would be expected at both signs of the applied bias.
On the other hand, if the normal state (pseudo)gap has no direct
   connection with the superconducting gap, as in the DDW scenario,
   then only a definite kind of peaks, either of electron or of hole
   kind, should appear.

The spatial inhomogeneity of the local electronic structure around an
   isolated impurity in an unconventional superconductor has been
   already studied \emph{via} the solution of the Bogoliubov-De~Gennes
   equations for a disordered system
   \cite{Ghosal:00,Haas:00,Maki:00,Haas:00a}, and within a
   self-consistent approach, taking into account the changes in the
   order parameter induced by the impurity potential \cite{Flatte:97}.
Here, we follow a different approach, and address the problem of
   finding the change in the local electron density around an isolated
   impurity in an anisotropic superconductor, within linear response
   theory.

The outline of the paper is then as follows.
After reviewing the expression of the linear response function $F$
   in real space for a uniform electron gas \cite{March:60}, in
   Sect.~\ref{sec:linear} we calculate the first-order Dirac density
   matrix, both for an isotropic, $s$-wave superconductor in three
   dimensions (3D), and for an anisotropic, $d$-wave superconductor in two
   dimensions (2D).
In Sect.~\ref{sec:F} we then generalize the above expression for $F$ to the
   superconducting case, but now in momentum space, where it is
   directly related to the Fourier transform of the electron density
   change in a superconductor around a highly localized impurity.
A numerical analysis then demonstrates that a $d$-wave order parameter
   gives rise to an azymuthal modulation in the momentum dependence of
   $F$, which is responsible of the four-lobed pattern observed in the
   density change around an impurity.
Later in Sect.~\ref{sec:conclusions} we summarize and give directions
   for future work.

\section{Linear response theory for normal and superconducting state}
\label{sec:linear}

The linear response function $F(\br,\br^\prime ,\EF )$ for a one-body
   potential $V(\br)$ introduced into an initially uniform Fermi gas
   provides the density change $\delta\rho(\br)$ in that gas due to
   the `impurity' generating $V(\br)$.
This quantity $F$ is given in the early work of March and
   Murray \cite{March:60} and is translationally invariant and
   spherically symmetric, namely
\begin{equation}
F(\br,\br^\prime ,\EF ) \equiv F(|\br - \br^\prime |, \EF) =  
- \frac{\kF^2}{2\pi^3}
   \frac{j_1 (2\kF |\br -\br^\prime |)}{|\br -\br^\prime |^2} ,
\label{eq:MM}
\end{equation}
where $j_1 (x) = (\sin x - x\cos x)/x^2$ is the spherical Bessel
   function of order one, $\kF$ the modulus of the Fermi momentum,
   and $\EF = \frac{1}{2}\kF^2$ the Fermi energy (in units such that
   $\hbar = m = 1$, to be used throughout the present work).
Here, prompted by the rekindling of interest in defect problems in
   superconductors \cite{Pan:99,Hudson:99,Hudson:01,Pan:01}, the
   analogous response function is calculated within 
   the BCS model of the superconducting phase \cite{Schrieffer:64},
   for both isotropic and anisotropic superconductors.
Due to the `rounding off' of the Fermi momentum distribution $n(\bp)$, 
   which at $T=0$ in the normal state is the usual step function
   terminating at Fermi momentum $p_{\rm F} = \hbar\kF$, the
   superconducting ($S$) response function $F_S (\br - \br^\prime ,\EF )$
   has more pronounced `damped' oscillations at large separations
   $|\br - \br^\prime |$, as compared to the normal state case, the
   damping factor of the oscillations measuring the `blurring' of the
   `edge' at $n(p_{\rm F} )$ due to Cooper pair formation.
In addition to that, for an anisotropic superconducting state, as is
   the case for a $d$-wave order parameter, the superconducting
   response function $F_S (\br - \br^\prime ,\EF)$ is expected to lose
   spherical symmetry.

Returning to the study of March and Murray \cite{March:60}, these
   authors obtained the displaced charge $\delta\rho(\br)$ due to a
   perturbation $V(\br)$ introduced
   into an initially uniform Fermi gas having electron density
   $\rho_0 = \kF^3 /3\pi^2$ as
\begin{equation}
\delta\rho(\br) = \int V(\br^\prime ) F(|\br
   -\br^\prime | ,\EF) \d^3 \br^\prime .
\label{eq:MMdrho}
\end{equation}
Here, our interest is to generalize 
   Eq.~(\ref{eq:MMdrho}) in order to treat the effect of a given perturbation
   $V(\br)$ introduced into the superconducting phase as described by
   the BCS model \cite{Schrieffer:64}.
Then we can write the analogue of Eq.~(\ref{eq:MMdrho}) in the form
\begin{equation}
\delta \rho_S (\br ) = \int V(\br^\prime ) F_S ( \br -\br^\prime ,\EF)
   \d^3 \br^\prime . 
\label{eq:Sdrho}
\end{equation}

To proceed to calculate $F_S$ below, we note next the result of
   Stoddart, March, and Stott \cite{Stoddart:69} (see also
   Ref.~\cite{Jones:73-2}) that
\begin{equation}
\frac{\partial F(\br,\br^\prime ,E)}{\partial E} = 2 \Re \left[
   G(\br,\br^\prime , E^+ ) \frac{\partial\gamma (\br,\br^\prime ,
   E)}{\partial E} \right],
\label{eq:Stoddart}
\end{equation}
where $E^+ = E + \iunit\delta$.
Inserting the Green function for an outgoing spherical wave
\begin{equation}
G(\br,\br^\prime ,E^+ ) = \frac{\exp (\iunit k |\br-\br^\prime |)}{4\pi
   |\br-\br^\prime |} ,
\label{eq:Gnormal}
\end{equation}
where $E=\frac{1}{2} k^2$, and the first-order Dirac density matrix
\begin{equation}
\gamma(\br,\br^\prime ,\EF) = 2 \sum_{|\bk| < \kF} \exp [\iunit
   \bk\cdot(\br - \br^\prime )] = \frac{\kF^3}{\pi^2} \frac{j_1 (\kF
   |\br - \br^\prime |)}{\kF |\br - \br^\prime |}
\label{eq:gammaN}
\end{equation}
into Eq.~(\ref{eq:Stoddart}), one regains the response function,
   Eq.~(\ref{eq:MM}), after some manipulation.
We also remind that Eq.~(\ref{eq:Stoddart}) can be equivalently cast
   in the integral form \cite{Flores:79}:
\begin{equation}
F(\br,\br^\prime ,\EF ) = i \int G(\br,\br^\prime,E) G(\br^\prime
   ,\br,E) \d E.
\label{eq:Flores}
\end{equation}

\begin{figure}[t]
\centering
\includegraphics[height=\columnwidth,angle=-90]{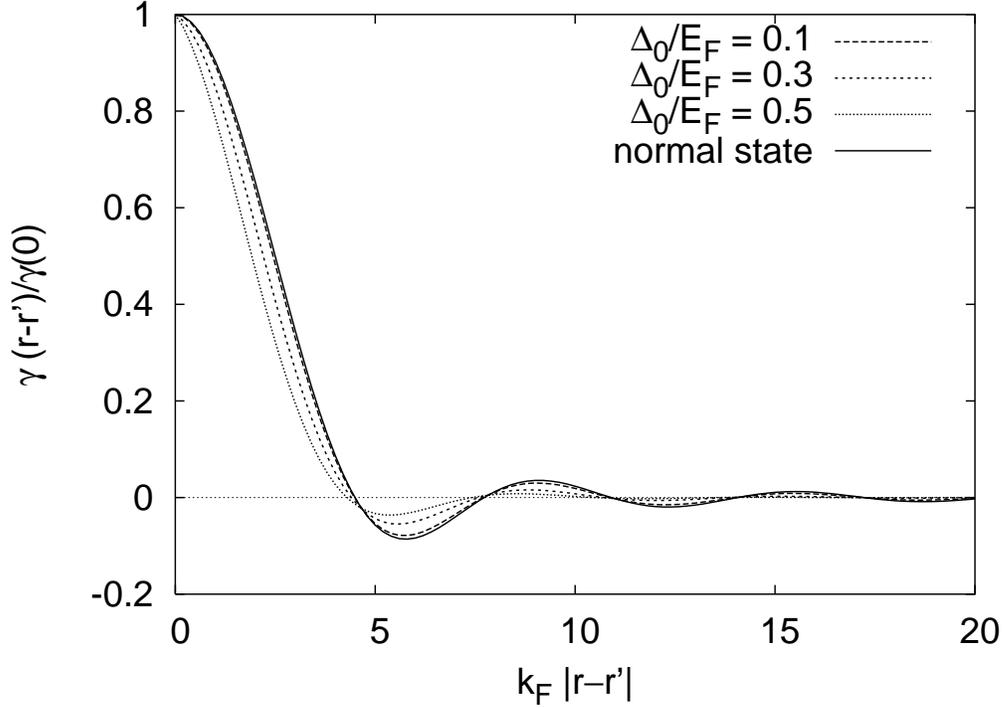}
\caption{Shows ratio of first-order density matrix to its diagonal
   value in 3D for three different values of the ratio $\Delta_0 /\EF =
   0.1$, $0.3$, $0.5$.
Solid line is normal state result.
It should be actually emphasized that $\Delta_0 /\EF < 1$ for most
   cuprates.
The larger values of the parameter $\Delta_0 /\EF$ used here and in
   Fig.~\protect\ref{fig:gammad} below are intended to magnify the
   effect of a superconducting gap in the first-order Dirac density
   matrix.
}
\label{fig:gamma}
\end{figure}

In the superconducting state, the upper branch of the Bogoliubov
   excitations is characterized by a dispersion relation $E_\bk =
   \sqrt{\xi_\bk^2 + \Delta_\bk^2}$, where $\xi_\bk = \frac{1}{2} (k^2
   - \kF^2 )$ is the free particle dispersion, measured with respect
   to the Fermi level, and $\Delta_\bk$ is the superconducting gap at
   $T=0$ (in general, a function of momentum $\bk$).
Eq.~(\ref{eq:gammaN}) is immediately generalized then to read in the
   superconducting phase
\begin{equation}
\gamma_S (\br,\br^\prime ,\EF) = 2\sum_\bk n_S (\bk) \exp[\iunit\bk\cdot
   (\br - \br^\prime )],
\label{eq:gammaS}
\end{equation}
where $n_S (\bk)$ denotes the momentum distribution in the
   superconducting phase at $T=0$, which coincides with the coherence
   factor $v^2_\bk = \frac{1}{2} (1-\xi_\bk /E_\bk )$ of BCS theory
   \cite{Schrieffer:64}. 
In the $s$-wave case, corresponding to having a constant gap
   $\Delta_\bk = \Delta_0$ over momentum space, $v^2_\bk$ is
   characterized by an inflection point at $\kF$ of width $\sim
   2\Delta_0$ (see Fig.~2.4 in Ref.~\cite{Schrieffer:64}, or Fig.~7.12
   in Ref.~\cite{March:67}), and $ \gamma_S (\br , \br^\prime ,\EF)
   \equiv \gamma_S (|\br - \br^\prime |,\EF)$.
We have plotted Eq.~(\ref{eq:gammaS}) for $\gamma_S (|\br - \br^\prime 
   |,\EF)/\gamma_S (0,\EF)$ as a function of separation $|\br - \br^\prime 
   |$ for three instances of the ratio $\Delta_0 /\EF$ in
   Fig.~\ref{fig:gamma}, where these three curves are compared with
   the normal state result, Eq.~(\ref{eq:gammaN}).
The `damping' of the oscillations in the normal state response
   function on going to the superconducting phase is evident from the
   plots in Fig.~\ref{fig:gamma}.

\begin{figure}[t]
\centering
\begin{minipage}[c]{0.45\columnwidth}
\includegraphics[bb=156 151 474 453,clip,width=\textwidth]{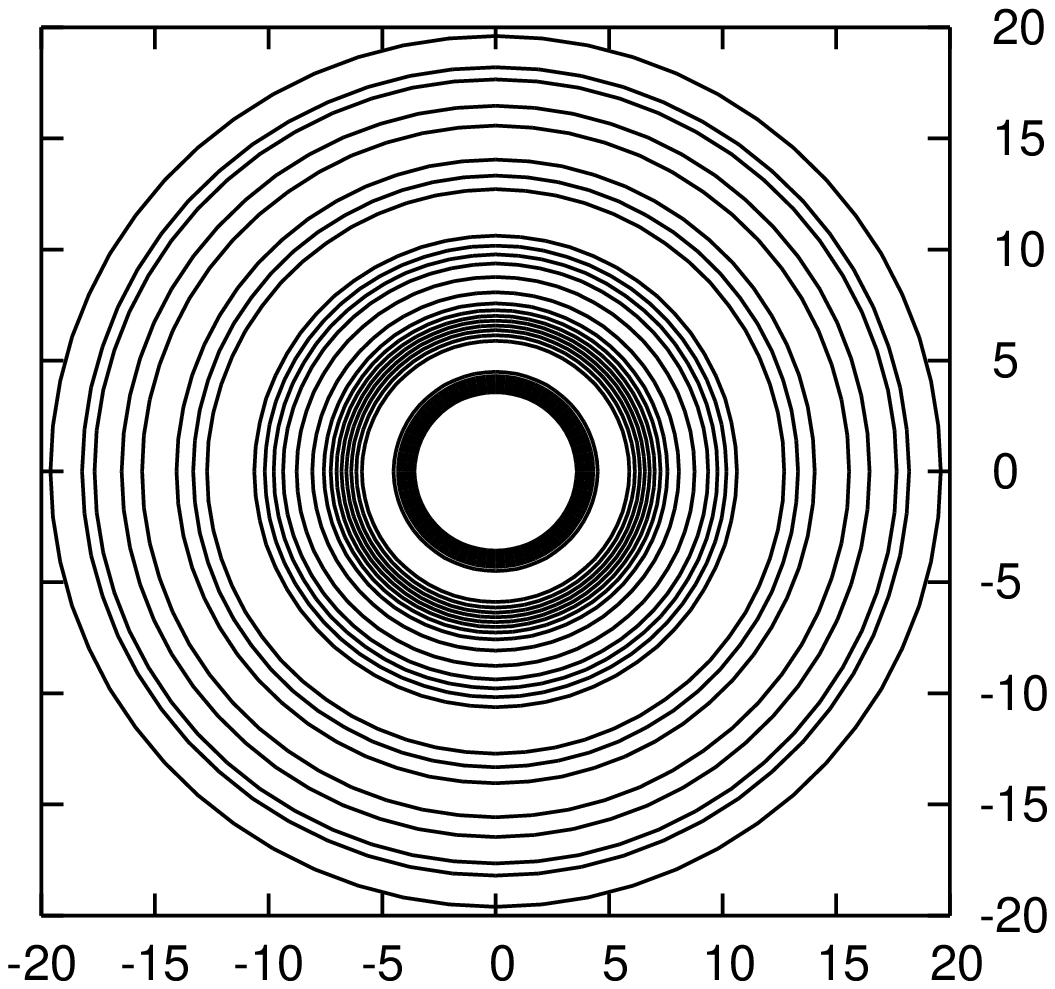}
\end{minipage}
\begin{minipage}[c]{0.45\columnwidth}
\includegraphics[bb=156 151 474 453,clip,width=\textwidth]{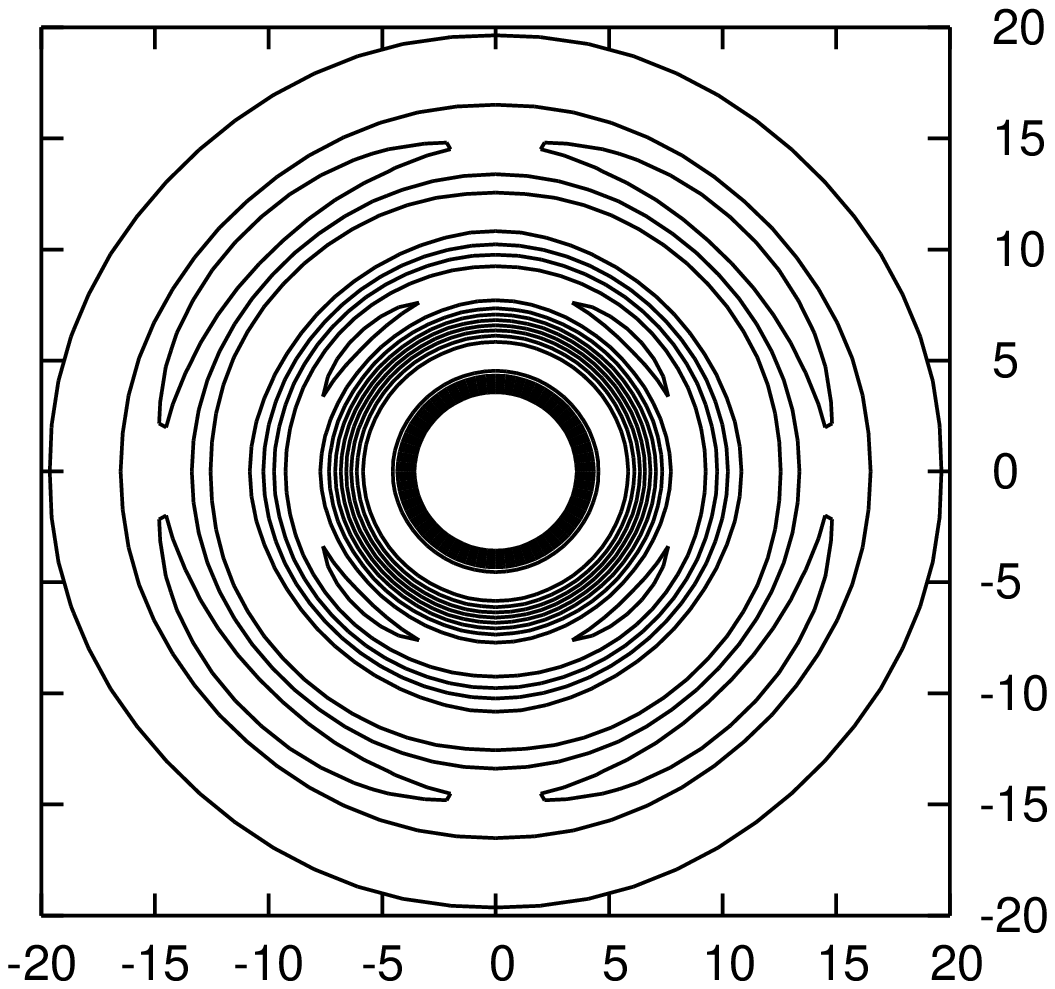}
\end{minipage}
\begin{minipage}[c]{0.45\columnwidth}
\includegraphics[bb=156 151 474 453,clip,width=\textwidth]{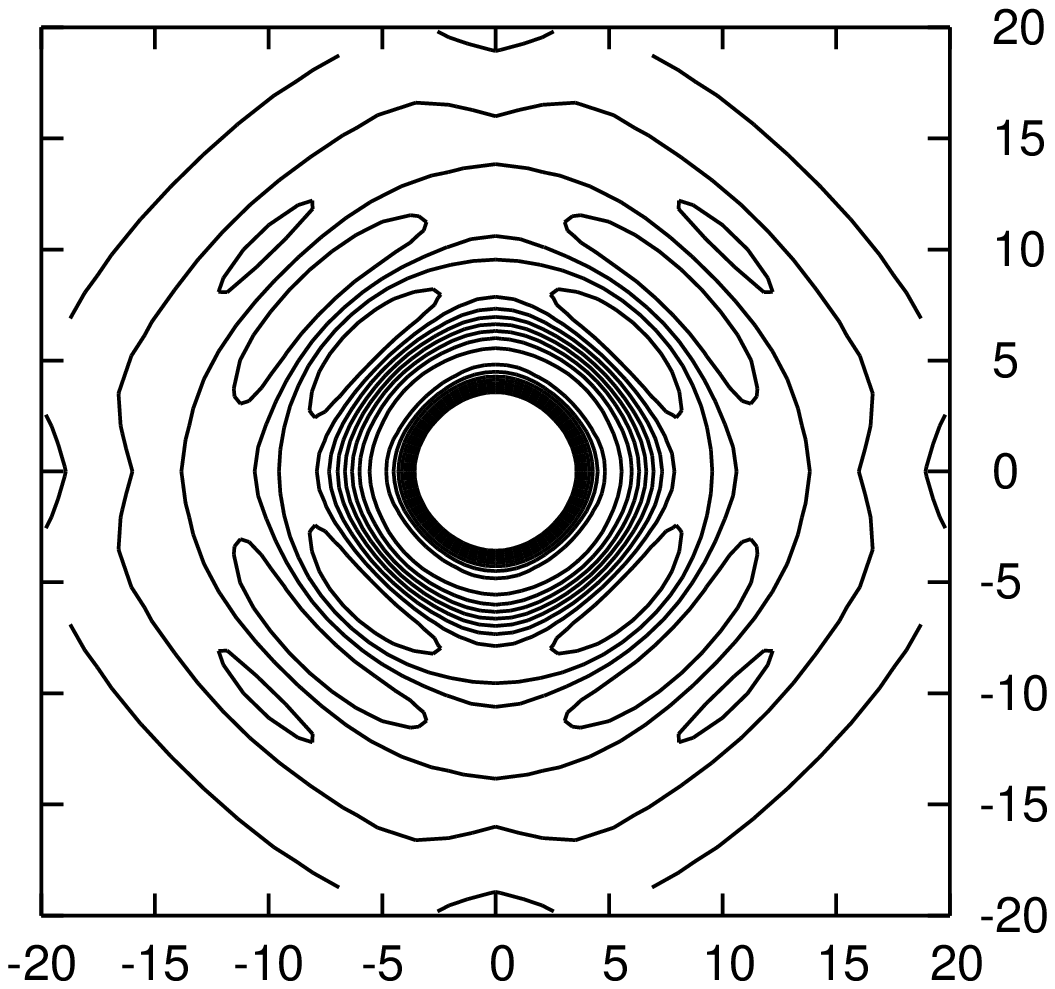}
\end{minipage}
\begin{minipage}[c]{0.45\columnwidth}
\includegraphics[bb=156 151 474 453,clip,width=\textwidth]{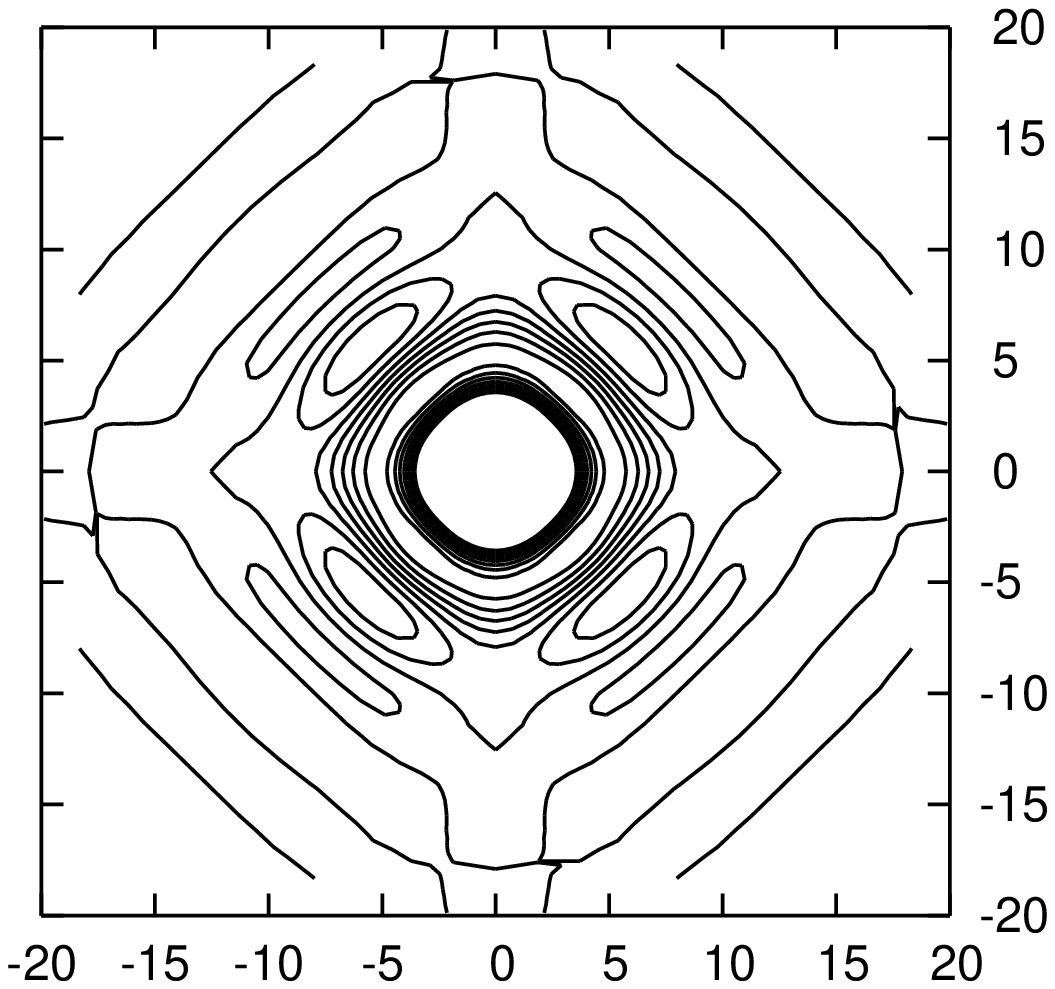}
\end{minipage}
\caption{Shows ratio of first-order density matrix to its diagonal
   value for three different values of the ratio $\Delta_0 /\EF = 0$
   (normal case), $0.1$, $0.3$, $0.5$ (left to right, bottom to top),
   but now for the 2D, $d$-wave case. 
As in Fig.~\protect\ref{fig:gamma}, all contour plots are against $\kF 
   ({\bf r} - {\bf r}^\prime )$ (now a 2D vector).
Damping of the normal-state oscillations is more pronounced in the
   fully gapped $x$ and $y$ directions than in the gapless $y=\pm x$
   directions.
}
\label{fig:gammad}
\end{figure}

The above derivation can be readily generalized to an anisotropic
   superconductor.
In the case of the high-$T_c$ cuprates, we assume a $d$-wave gap
   $\Delta_\bk = \Delta_0 \cos 2\theta$, with $\bk$ now a
   two-dimensional (2D) wavevector, $\theta$ being the angle it forms
   with the $x$ direction.
Such an order parameter is characterized by nodes for
   $\theta=\pm\pi/2$, corresponding to the directions $y=\pm x$.
In the normal case, but now in 2D, we find the translationally
   invariant, spherically symmetric first-order Dirac density matrix:
\begin{equation}
\gamma(\br,\br^\prime ,\EF) = \frac{\kF^2}{\pi} \frac{J_1 (\kF |\br -
   \br^\prime |)}{\kF |\br - \br^\prime |} ,
\end{equation}
with $J_1 (x)$ here denoting the Bessel function of the first kind and
   order one, which again displays oscillations as
   Eq.~(\ref{eq:gammaN}), but now with a weaker asymptotic decay in
   the limit $\kF |\br - \br^\prime | \gg 1$ (see
   Fig.~\ref{fig:gammad}, upper left panel).
In the superconducting case, taking into account the $d$-wave symmetry
   of the order parameter, $\gamma_S (\br-\br^\prime , \EF)$,
   Eq.~(\ref{eq:gammaS}), loses its spherical symmetry.
We plot $\gamma_S (\br-\br^\prime ,\EF)/\gamma_S (0,\EF)$ in
   Fig.~\ref{fig:gammad} as a function of the 2D vector
   $\br-\br^\prime$.
As in the isotropic, 3D case, we recover a `damping' of the
   oscillations in the normal state response function in going into
   the superconducting phase, which is now more pronounced in the
   fully gapped $x$ and $y$ directions, than in the gapless directions
   $y=\pm x$.

\section{Linear response function for an anisotropic superconductor}
\label{sec:F}

We now turn to the problem of explicitly evaluating the linear
   response function for a 2D superconductor, both in the $s$- and in
   the $d$-wave case.
We start by noting that Eqs.~(\ref{eq:MMdrho}--\ref{eq:Sdrho}) express
   the relation between the density 
   change $\delta\rho$ and the impurity potential $V$ as a convolution
   in real space, the convolution kernel being the linear response
   function $F$.
Such a convolution naturally translates into a simple product in
   momentum space, $\delta\rho (\bq) = V(\bq) F_S (\bq,\EF)$.
Moreover, in the case of a highly localized impurity, as is the case
   of a $\delta$-function-like impurity potential in real space
   \cite{Byers:93}, one has $V(\bq)=V_0$, so that the Fourier
   transform of the linear response function $F_S (\bq,\EF)$
   practically coincides with the Fourier transform of the density
   change.
In the normal state, the latter quantity is readily obtained by
   Fourier transforming Eq.~(\ref{eq:Flores}) as a convolution of
   Green's functions in momentum space.
In the superconducting state, such an expression is generalized by the
   static limit ($\omega=0$) of the polarization function,
   which for a 2D superconductor reads \cite{Prange:63} 
\begin{eqnarray}
F_S (\bq,\EF) &=& \frac{i}{2\pi} \Tr \int\d E \int \frac{\d^2 \bk}{(2\pi)^2}
   \btau_3 \bG (\bk,E) \btau_3 \bG (\bk-\bq,E) \nonumber\\
&=& 2 \int \frac{\d^2 \bk}{(2\pi)^2} \frac{\left( u_{\bk+\bq} v_{\bk} + u_\bk
   v_{\bk+\bq} \right)^2}{E_{\bk+\bq} + E_\bk} ,
\label{eq:Prange}
\end{eqnarray}
where again $v^2_\bk$ and $u^2_\bk = 1-v^2_\bk$ are the coherence
   factors of BCS theory \cite{Schrieffer:64}, but now in general
   depending on an anisotropic gap $\Delta_\bk$.
In Eq.~(\ref{eq:Prange}), $\bG$ denotes the matrix Green's function in
   Nambu notation \cite{Schrieffer:64}, and $\btau_3$ is a Pauli matrix.

\begin{figure}[t]
\centering
\includegraphics[height=\columnwidth,angle=-90]{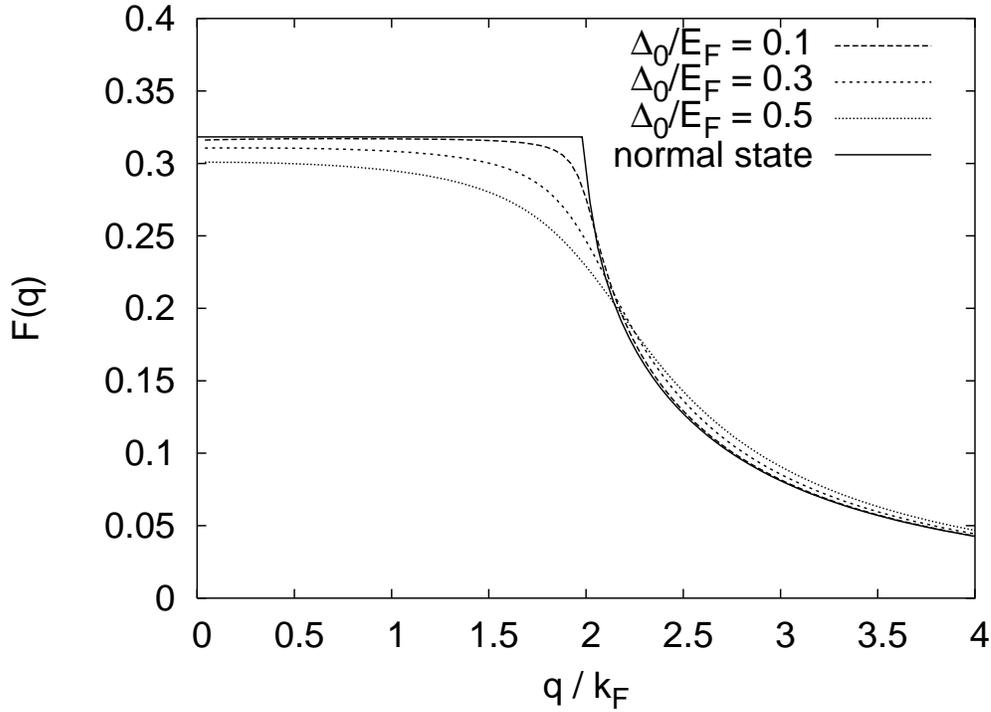}
\caption{Linear response function in
   momentum space, $F_S (q,\EF)$, Eq.~(\protect\ref{eq:Prange}), in the
   case of a translationally invariant, isotropic 2D system.
Dashed line shows the normal state analytical result,
   Eq.~(\protect\ref{eq:Fnormal}).
Continuous lines show our numerical results for the superconducting
   state, when an $s$-wave order parameter is assumed, for the same
   values of $\Delta_0 /\EF$ as in Fig.~\protect\ref{fig:gamma}.
Notice the step discontinuity at $q=2\kF$ in the normal state, which
   gets smoothed over a width $\sim 2\Delta_0$ in the superconducting
   case.
}
\label{fig:Fs}
\end{figure}

In the normal state ($\Delta_0 \to 0$), the integration in
   Eq.~(\ref{eq:Prange}) is feasible analytically, and we find:
\begin{equation}
F(q,\EF) \equiv F_S (\bq,\EF;\Delta_0 = 0) = \left\{
\begin{array}{ll}
\displaystyle\frac{1}{\pi} , & \quad \mbox{if $q\leq 2\kF$,}\\
\displaystyle\frac{1}{\pi} \frac{q-\sqrt{q^2 -4}}{q} , & \quad \mbox{if $q > 2\kF$.}
\end{array}\right.
\label{eq:Fnormal}
\end{equation}
Therefore, we find that the linear 
   response function for a 2D system in the normal state is
   characterized by a step discontinuity at $q=2\kF$
   (Fig.~\ref{fig:Fs}), which is responsible of damped Friedel-like
   oscillations of period $\sim \pi/\kF$ for $F(\br,\EF)$ in real
   space.

In the superconducting state, for an $s$-wave, isotropic order
   parameter in 2D, such a discontinuity is smoothed over a width of
   $\sim 2\Delta_0$ around $q=2\kF$, as is shown numerically in
   Fig.~\ref{fig:Fs}.
This is again expected to give rise to more pronounced damped
   oscillations in the $r$-dependence of the linear response function
   in real space.

\begin{figure}[t]
\centering
\begin{minipage}[c]{0.45\columnwidth}
\includegraphics[bb=62 56 541 528,clip,width=\textwidth]{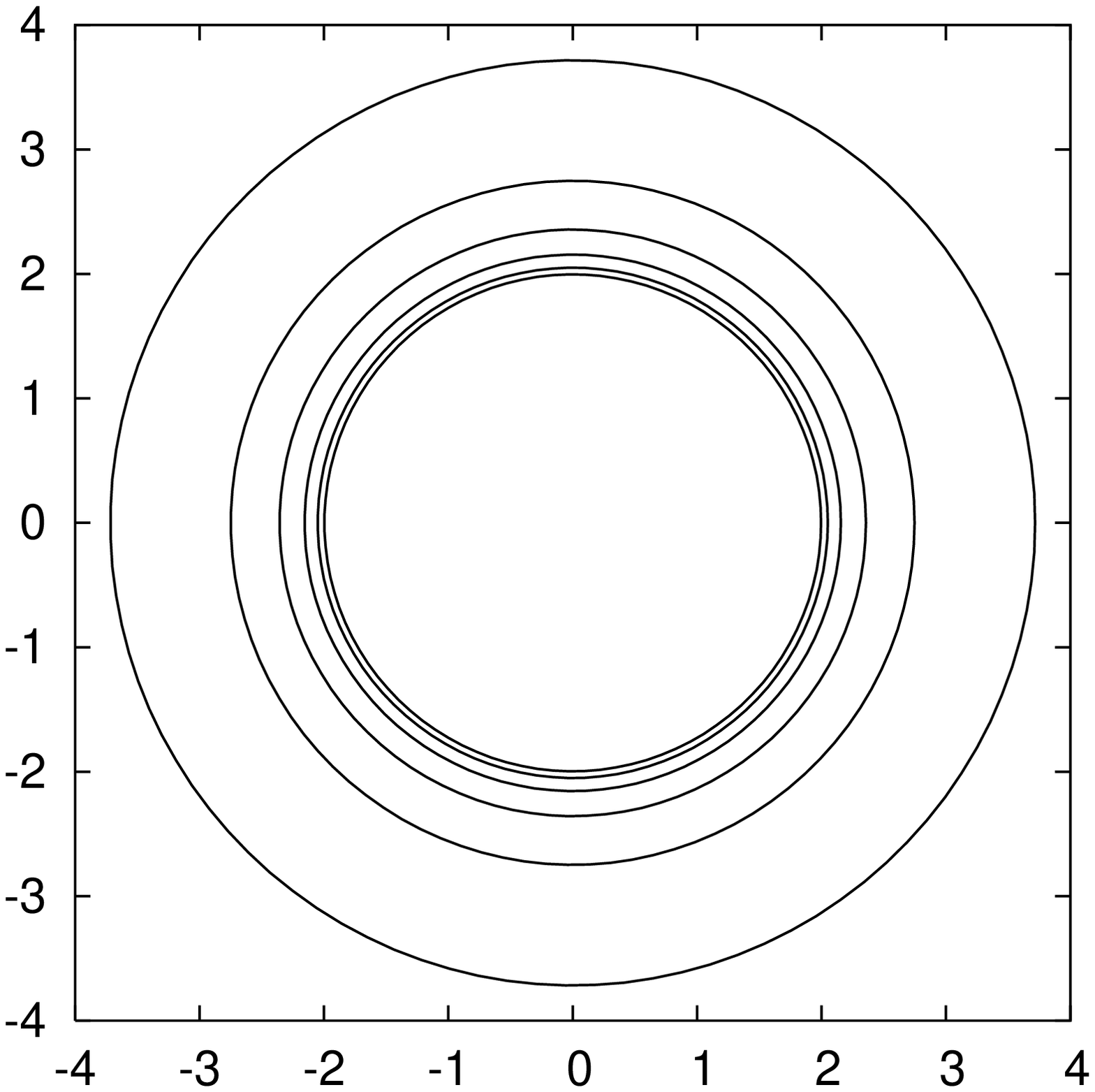}
\end{minipage}
\begin{minipage}[c]{0.45\columnwidth}
\includegraphics[bb=62 56 541 528,clip,width=\textwidth]{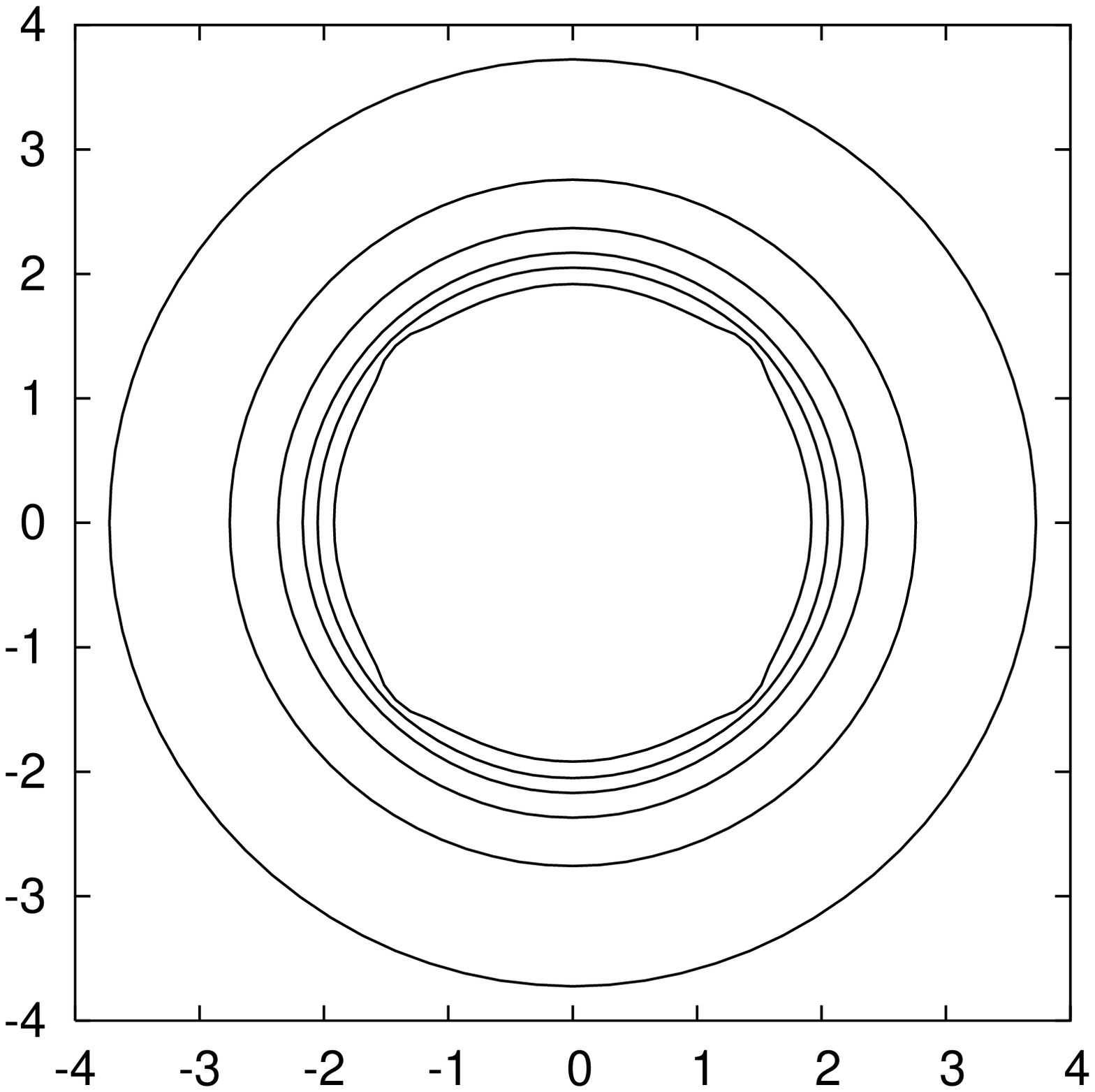}
\end{minipage}
\begin{minipage}[c]{0.45\columnwidth}
\includegraphics[bb=62 56 541 528,clip,width=\textwidth]{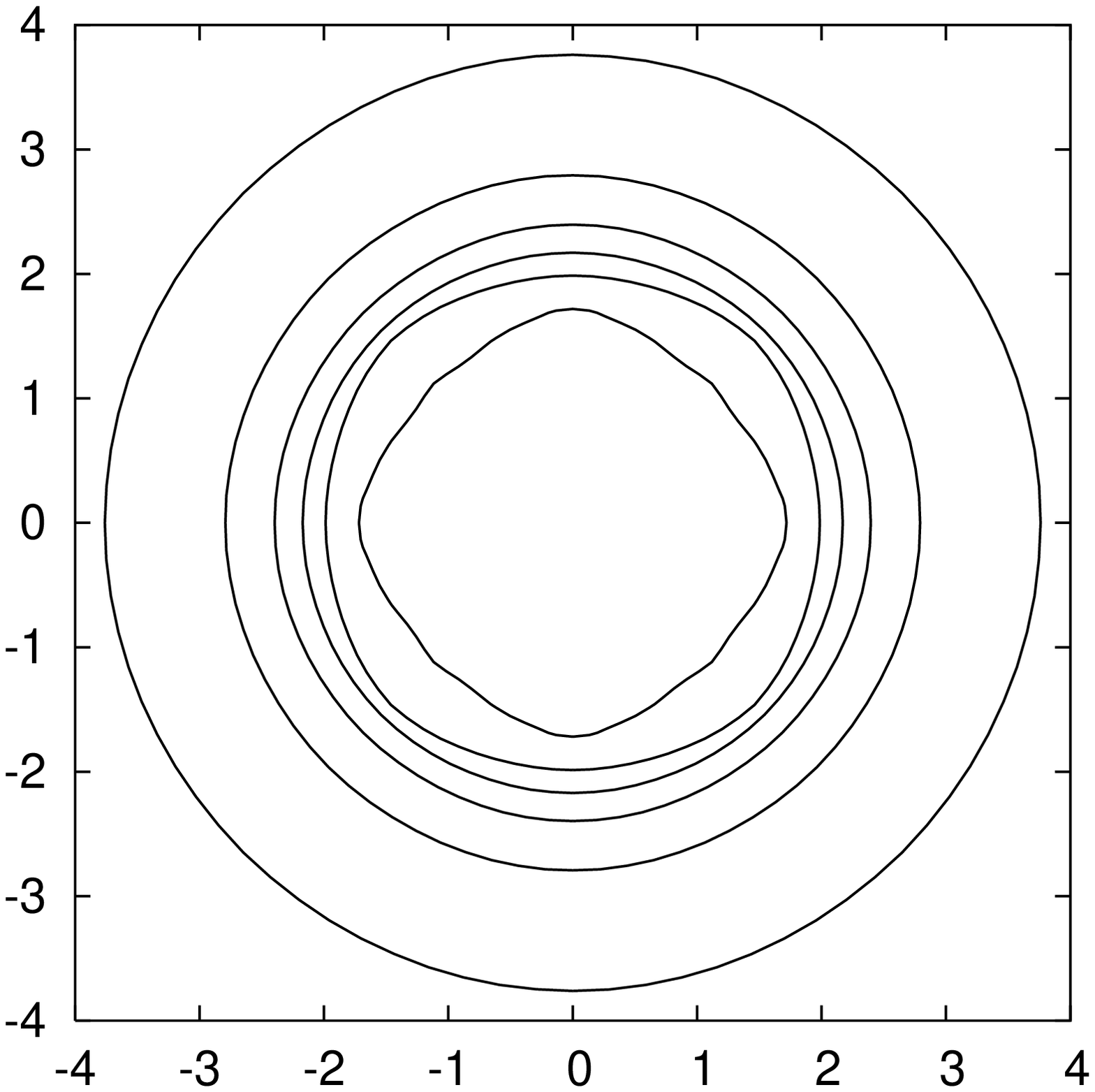}
\end{minipage}
\begin{minipage}[c]{0.45\columnwidth}
\includegraphics[bb=62 56 541 528,clip,width=\textwidth]{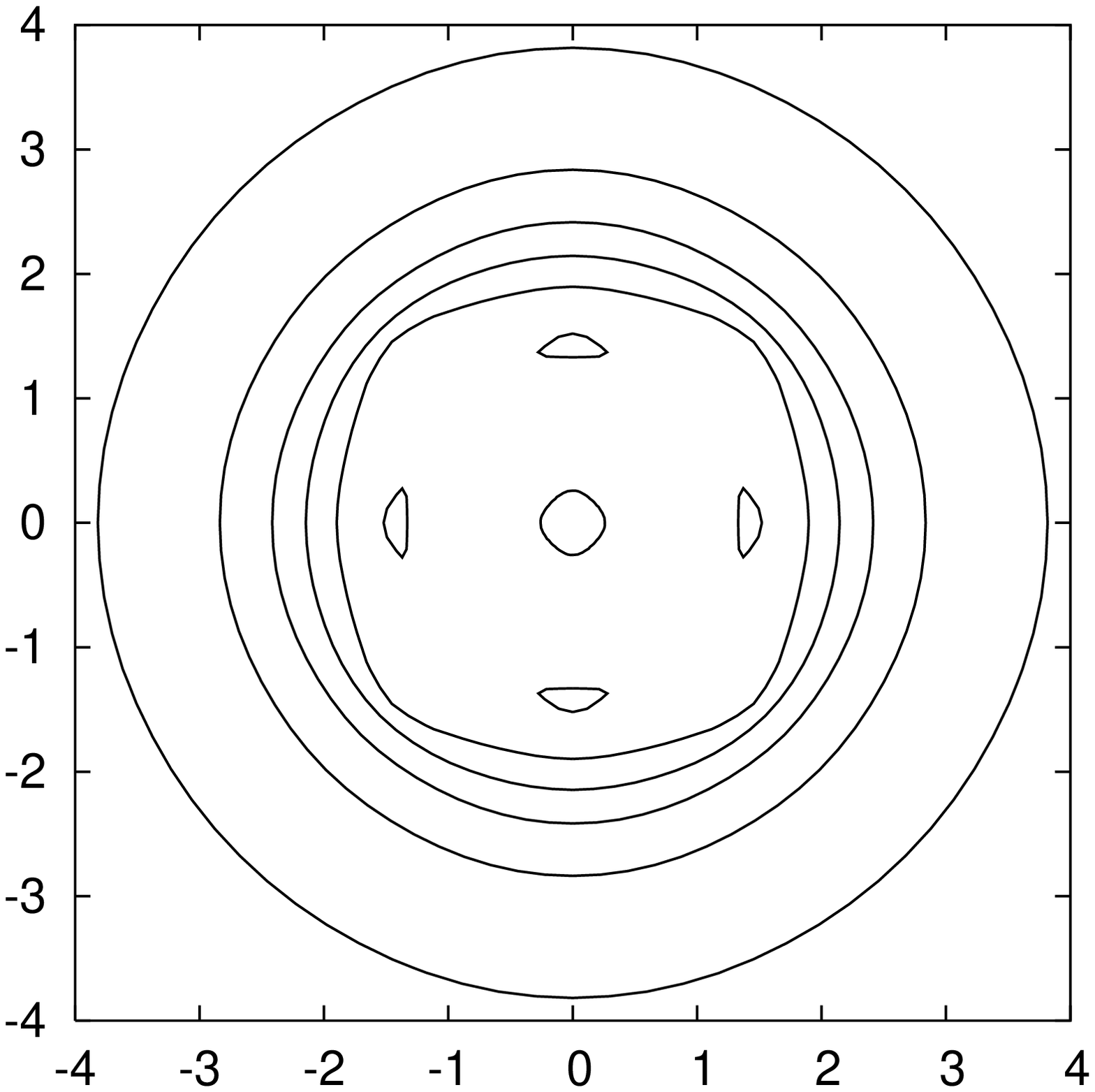}
\end{minipage}
\caption{Linear response function $F_S (\bq,\EF)$,
   Eq.~(\protect\ref{eq:Prange}), for a $d$-wave superconductor in 2D.
Actually displayed are contour plots of $F_S (\bq,\EF)$ as a function
   of $\bq / \kF$ for $\Delta_0 /\EF = 0.0$ (normal state), $0.1$,
   $0.3$, $0.5$ (left to right, top to bottom).
In going into the superconducting state, the step discontinuity at
   $q=2\kF$ gets smoothed more in the $q_x$ and $q_y$ directions
   (corresponding to gap maxima) than in the $q_y = \pm q_x$
   directions (corresponding to gap nodes).
}
\label{fig:Fd}
\end{figure}

In the case of a 2D, $d$-wave superconductor, $F_S (\bq,\EF)$ must be
   analyzed as a function of $\bq$ as a vector.
Numerical results are plotted in Fig.~\ref{fig:Fd} for different
   values of the ratio $\Delta_0 /\EF$, and compared to the
   (isotropic) result for the normal state. 
As in the isotropic, $s$-wave case, the opening of a superconducting
   gap at the Fermi level tends to smear out the step discontinuity at
   $q=2\kF$ over a width $\sim 2\Delta_0$.
However, such an effect is more enhanced in the $q_x$ and $q_y$ directions,
   corresponding to maxima in the energy gap, than in the $q_y = \pm
   q_x$ directions (gap nodes), where it is virtually absent.
This gives rise to an azymuthal modulation of $F_S (\bq,\EF)$, of period
   $\pi/2$, which is ultimately responsible for the clover pattern in
   the density change around an isolated impurity, as observed
   in STM imaging experiments.

\section{Conclusions and directions for future work}
\label{sec:conclusions}

Motivated by the recent imaging results around an isolated impurity in
   HTSC
   \cite{Pan:99,Hudson:99,Hudson:01,Pan:01,Iavarone:02,Salluzzo:01},
   we have analyzed the linear response function in momentum space for
   a 2D uniform electron gas, both in the normal and in the superconducting
   state.
The opening of an energy gap at the Fermi level manifests itself
   already in the radial dependence of the first-order Dirac density
   matrix.
In the $s$-wave case, its diagonal element $\gamma_S (|\br-\br^\prime
   |,\EF)$ decays more rapidly with distance from the diagonal than
   does its normal state counterpart. 
In the $d$-wave case, fingerprints of the anisotropic momentum
   dependence of the energy gap are already present in the angular
   dependence of $\gamma_S (\br-\br^\prime ,\EF)$, with more
   pronounced damping of the oscillations in the fully gapped $x$ and
   $y$ directions than in the gapless $y=\pm x$ directions.
Such a behaviour is confirmed by the momentum dependence of the linear
   response function $F_S (\bq,\EF)$.
In the normal, isotropic state, the latter function is characterized
   by a step discontinuity at $q=2\kF$, giving rise to (damped)
   Friedel-like oscillations in its real space counterpart.
The discontinuity at $q=2\kF$ gets smoothed over a width $\sim
   2\Delta_0$ in going into the superconducting state.
For a $d$-wave superconductor, however, such an effect is more
   pronounced in the fully gapped $q_x$ and $q_y$ directions, than
   along the $q_y = \pm q_x$ nodes.
This gives rise to an azymuthal modulation of $F_S (\bq,\EF)$, which
   is ultimately responsible of the four-lobed pattern in the density
   change, observed in STM imaging experiments around an impurity.
Before a direct comparison could be made with such experimental
   results, however, we believe that a 
   more accurate band dispersion relation should be considered (see,
   \emph{e.g.,} Ref.~\cite{Martin:00a}), as
   well as a more realistic choice for the impurity potential, which
   we reserve to future work.

\begin{ack}
The authors are indebted with M.~Salluzzo, F.~Siringo, and J. Verbeeck
   for helpful discussions.
One of us (N.~H.~M.) made his contribution to this work during a visit 
   to the Physics Department, University of Catania, in the year 2000.
Thanks are due to the Department for the stimulating environment and
   for much hospitality.
N.~H.~M. also wishes to thank Professor V.~E. Van Doren for his
   continuing interest and support.
G.~G.~N.~A. acknowledges the University of Antwerp (RUCA) for
   much hospitality during the period in which this work was revised
   and brought to completion.
\end{ack}

\bibliography{a,b,c,d,e,f,g,h,i,j,k,l,m,n,o,p,q,r,s,t,u,v,w,x,y,z,zzproceedings,Angilella}
\bibliographystyle{elsart-num}

\end{document}